\RequirePackage{tikz}
\documentclass[pdflatex,sn-mathsphys]{snA4}

\newif\ifcolour
\colourtrue

\usepackage{hyperref}
\usetikzlibrary{calc}
\usepackage{amsmath}
\usepackage{amssymb}
\usepackage{commath}
\usepackage{gensymb}
\usepackage{graphicx}
\usepackage{caption}
\usepackage{subcaption}
\usepackage{multirow}
\usepackage{cleveref}

\newcommand{\dB}{\,\mathrm{dB}}
\newcommand{\Hz}{\,\mathrm{Hz}}
\newcommand{\kHz}{\,\mathrm{kHz}}
\newcommand{\ms}{\,\mathrm{ms}}

\newcommand{\figscale}{0.4}
\newcommand{\smallfigscale}{0.35}

\newcommand{\figlegendscale}{0.8}

\ifcolour
\graphicspath{{Figures/} {Figures/freq_range_figs/}}
\else
\graphicspath{{Figures/monochrome_figures/} {Figures/}}
\fi

\begin{document}
	
	\author{\fnm{Keziah} \sur{Milligan}}
	\author{\fnm{Nicholas~J.} \sur{Bailey}}\email{%
          nick@n-ism.org}
        \author{\fnm{Bernd} \sur{Porr}}\email{%
          bernd.porr@glasgow.ac.uk}     
    \affil{\orgdiv{James Watt School of Engineering},
               \orgname{The University of Glasgow},
               \orgaddress{\city{Glasgow}, \postcode{G12 8LT}, \country{United Kingdom}}}
  
	\title[The Time and Frequency Response of Hopf Resonators]{An Empirical Investigation Into the Time and Frequency Response Characteristics 
	of Hopf Resonators}
	\maketitle

	\abstract{%
	We present an empirical investigation of software developed 
	by the Science and Music Research Group at the University of Glasgow. \footnote{The complete 
	source code for this project can be found at 
	\url{https://github.com/keziah55/DetectorBank}.} Initially created for musicological 
	applications, it is equally applicable in any area where precise time and frequency 
	information is required from a signal, without encountering the problems associated with 
	the uncertainty principle.
	By constructing a bank of non-linear tuned resonators (`detectors'), each of which 
	operates at a Hopf bifurcation, it is possible to detect frequencies within half a 
	period of oscillation, even in the presence of wideband noise. 
	
	The time and frequency response characteristics of these detectors will be examined 
	here.
	%
    }


	\section{Introduction}
	\label{sec:intro}
	
	There are diverse engineering applications of the rapid and accurate
	detection of the onset of a frequency component hidden within a
	wideband signal.
	A change in the spectrum of a stator current of an induction motor \cite{2000benbouzid_induction_motor_sig} which is about to fail;
	time-frequency analysis of mechanical systems to detect faults 
	\cite{2013feng_time_frequency_machinery_fault_detection}; extraction of anthropogenic climate 
	change from natural cyclic variations \cite{1996santer_climate_change_detection}:
	all of these activities depend on accurate detection of the onset
	of a particular spectral feature in the presence of significant
	energy in possibly very nearby parts of the spectrum. 
	
	There is an exemplar of this type of problem, for which accurate onset times
	of frequencies within a very narrow bandwidth is a prerequisite,
	in polyphonic (multiply simultaneously-sounding) note onset detection for music.
	This is a widely researched question which is so far only incompletely
	answered \cite{2005bello_onsets}. 
	Classical signal processing 
	methods having only partially succeeded \cite{2012bock_onset_detector}, more recent 
	research \cite{2005collins_comparison} has been directed towards the acknowledgement
	of psychoacoustic involvement in music perception and cognition.
	This is clearly justified insofar as even musically na\"\i{}ve
	listeners can easily outperform state-of-the-art algorithms in
	note-onset detection tasks. However, it does overlook the possibility
	of applying non-linear detection methods. These appear promising,
	as we know that the cochlea maps the frequency spectrum physically 
	\cite{2012moore_psychology_hearing},
	that it contains (in mammals) non-linear structures appearing to have
	the purpose of enhancing frequency selectivity \cite{2008hudspeth_ear_amplification}, 
	and that such mapping
	is physiological rather than psychoacoustic.
	We therefore present a non-linear
	detector which is highly frequency-specific, and is inspired by the
	ear's non-linear elements, although it in no way claims to simulate them
	(such models can be found elsewhere in the literature).
	The behaviour of non-linear systems is difficult to define when
	driven by broad-band signals, so we commence by empirically examining the
	behaviour of our candidate system as thoroughly as possible. We then
	present a software implementation of the model which is useful in
	determining onsets of narrow-band stimuli within a wideband signal.

	\subsection{Entropic uncertainty}
	\label{sec:uncertainty}
	
	Uncertainty principles refer to the impossibility having sharply localised representations of a 
	pair of properties of a function simultaneously \cite{2013ricaud_uncertainty}. In signal 
	processing, the properties which cannot be simultaneously measured are time and frequency. 
	To have a frequency, signal must necessarily repeat. As it is impossible to refer to frequency 
	until the repetition is established, it is not meaningful to talk about frequency at an instant 
	in time. 
	It would, therefore, seem that there is a fundamental constraint that will prevent any 
	attempts to determine what frequencies are present in a signal and at exactly what time those 
	frequencies appear. \cite{1946gabor_communication} gives the relation between time and 
	frequency resolution, $\Delta t$ and $\Delta f$, as $\Delta t \Delta f \geq 0.5$.
	However, the human auditory system is not subject to these limitations 
	\cite{2018majka_ultrashort_acoustic_pulses}.
	
	By creating software inspired by the mechanical processes in the inner ear, we can 
	sidestep the issues that arise due to uncertainty. 
	We do not seek to measure the signal directly; instead the signal is 
	used to drive a bank of tuned resonators. Frequency and time information can then be 
	found from observable internal state variables,
	circumventing to a significant extent the entropic uncertainty
	arising with Fourier-based methods.

	\subsection{Auditory system}
	\label{sec:auditory_system}
	
	Time and pitch perception of sounds occurs in the cochlea.
	Incoming sound pressure waves cause travelling waves to appear
	across the basilar membrane \cite{2012olson_vonbekesy}.
	The point of maximum displacement along the membrane is proportional to the frequency of the sound, with high frequencies represented at the base of the membrane and low frequencies at the apex.
	However, this tonotopic map does not account for the
	pitch-acuity of human hearing \cite{2012plack_theories_of_hearing}.
	
	In the mammalian auditory system, the outer hair cells in the cochlea operate at a 
	Hopf bifurcation \cite{2000eguiluz_nonlinearities_in_hearing}. This amplifies, 
	compresses and 
	tunes the response of the basilar membrane \cite{2010hudspeth_hopf_cochlea} on the 
	stable side of the bifurcation and causes otoacoustic emissions 
	\cite{1978kemp_stimulated_otoacoustic_emissions,1979kemp_spontaneous_otoacoustic_emissions}
	 on the unstable side.

	\section{Hopf bifurcation}
	\label{sec:hopf}
	
	
	The equation for the Hopf bifurcation is a first order differential, with a control parameter, 
	$\mu$, which determines how close the system is to the bifurcation point, $\mu_0$. 	
	As the control parameter approaches and then crosses this critical value, the 
	conjugate 
	eigenvalues become purely imaginary, the system goes from stability to instability 
	and periodic orbits appear around an equilibrium.
	
	\subsection{Derivation of Hopf normal form}
	\label{sec:derivation_of_hopf}
	
	Following the procedures of Wiggins \cite{1990wiggins_nonlinear_dynamical_systems_chaos} and Kuznetsov \cite{2004kuznetsov_applied_bifurcation_theory}, the system of equations
	\begin{subequations}
		\begin{align}
		\dot{x_1} &= \mu x_1 - \omega x_2 - x_1(x_1^2 + x_2^2) \\
		\dot{x_2} &= \omega x_1 + \mu  x_2 - x_2(x_1^2 + x_2^2)
		\end{align}
		\label{eq:x_dot_system}
	\end{subequations}
	can be expressed as
	\begin{equation}
	\mathbf{f}(\mathbf{x}) = \mathbf{\dot{x}} = \mathbf{A} \mathbf{x} - (x_1^2 + x_2^2) \mathbf{x}
	\label{eq:x_dot_vectors}
	\end{equation}
	where $\mathbf{A}$ is the Jacobian of the system at equilibrium:
	
	\begin{equation}
	\mathbf{A} = \frac{\dif \mathbf{f}}{\dif \mathbf{x}} = \begin{bmatrix}
	\mu - 3x_1^2 - x_2^2 & -\omega - 2 x_1 x_2 \\
	\omega  - 2x_1 x_2 & \mu - x_1^2 - 3x_2^2
	\end{bmatrix}
	\label{eq:solve_jacobian_partial_derivatives}
	\end{equation}
	
	At equilibrium, $x_1 = x_2 = 0$, therefore
	\begin{equation}
	\mathbf{A} = \begin{bmatrix}
	\mu & -\omega \\
	\omega & \mu
	\end{bmatrix}
	\label{eq:jacobian_A}
	\end{equation}
	
	Observing the eigenvalues of the system determines not only the stability of the system, but also the frequency of the periodic solutions. 
	As $\mu$ tends to $\mu_0$, the angular frequency of the orbits tends to $\omega_0$. 
	Solving to find the eigenvalues of \eqref{eq:jacobian_A} gives
	\begin{equation}
	\lambda_{1,2} = \mu \pm j \omega
	\label{eq:eigenvalues}
	\end{equation}
	
	To find the normal form of the Hopf bifurcation,
	a complex variable is introduced:
	\begin{equation}
	z = x_1 + j x_2
	\label{eq:z=x1_plus_jx2}
	\end{equation} 
	So
	\begin{subequations}
	\begin{align}
	z^* &= x_1 - j x_2 \\
	zz^* &= \vert z\rvert^2 = x_1^2 + x_2^2 \\
	\dot{z} &= \dot{x}_1 + j \dot{x}_2
	\end{align}
	\label{eq:z_equations}
	\end{subequations}
	Combining \cref{eq:x_dot_system,eq:z=x1_plus_jx2,eq:z_equations} gives the normal 
	form of the Hopf bifurcation
	\begin{equation}
	\dot{z} = (\mu + j\omega)z - z\lvert z\rvert^2, \; z \in \mathbb{C}
	\label{eq:hopf_equation_normal_form_-1}
	\end{equation} 
	where the positive eigenvalue \eqref{eq:eigenvalues} appears as the coefficient of $z$. The 
	real part of the coefficient of the cubic term determines the stability of the orbits.
	
    The (real-valued) coefficient of $z|z|^2$ is known as the first Lyapunov coefficient,
    which will here be denoted 
	$b$. $b=-1$ in the above equation. The sign of $b$ determines 
	the stability of the solutions \cite{2008guckenheimer_hopf_bifurcations}. If  $b < 
	0$, the bifurcation is supercritical and the periodic orbits are stable; for $b > 0$ 
	there is a subcritical bifurcation and the orbits are unstable, therefore 
	\eqref{eq:hopf_equation_normal_form_-1} represents a supercritical Hopf bifurcation.
	
	
	Explicitly including the first Lyapunov coefficient gives a new form of the Hopf 
	bifurcation
	\begin{equation}
	\dot{z} = (\mu + j\omega) z + b \lvert z\rvert^2 z, \; z \in \mathbb{C}
	\label{eq:hopf_equation_normal_form_b}
	\end{equation}

	\subsection{Periodically forced Hopf bifurcation}
	\label{sec:forced_hopf}
	In order to use the Hopf bifurcation to analyse a signal, there must be an input, 
	which here takes the form of a sinusoidal forcing function $F(t)$.
	
	For a Hopf resonator at the bifurcation point, $\mu=\mu_0$, tuned to a characteristic 
	frequency $\omega_0$ with forcing frequency $\omega_{\textrm{in}}$ 
	\cite{2011zhang_periodically_forced_hopf_bifurcation}
	\begin{equation}
	\dot{z} = (\mu_0 + j \omega_0) z + b \lvert z\rvert^2 z + F(t)
	\label{eq:hopf_equation_forced}
	\end{equation}
	where $F = Xe^{j \omega_{\textrm{in}} t}$ or, in a system with real inputs, 
	$F = X\cos\,\omega_{\textrm{in}} t$.
	
\begin{figure*}
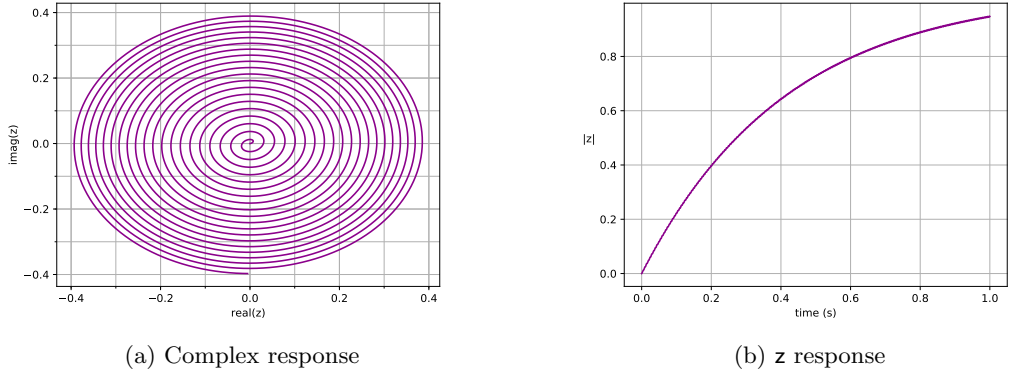

		\begin{subfigure}{0.49\textwidth}
			\centering
			\includegraphics[scale=\figscale]{100Hz_spiral_1.pdf}
			\caption{Complex response}
			\label{fig:forcing_example_100Hz_spiral}
		\end{subfigure}
		\begin{subfigure}{0.49\textwidth}
			\centering
			\includegraphics[scale=\figscale]{100Hz_line_1.pdf}
			\caption{$|z|$ response}
			\label{fig:forcing_example_100Hz_line}
		\end{subfigure}
		\caption{Output of Hopf system with characteristic and forcing frequencies both 
			at $100\Hz$. (a) first 200\,ms of the response in the complex plane, (b) 
			$|z|$ 
			response over 1 second.}
		\label{fig:forcing_example}
	\end{figure*}

	If the forcing frequency $\omega_{\textrm{in}}$ is equal to the characteristic frequency of 
	the system $\omega_0$, periodic orbits with angular frequency $\omega_0$ will appear. For 
	example, Figure \ref{fig:forcing_example} shows the response of a forced system with 
	a real input, where $\omega_0 = \omega_{\textrm{in}} = 200\pi\,\mathrm{rad\;s^{-1}}$ and 
	$X=25$. 
	(a) shows the first $200\ms$ of the response in the complex plane as the periodic 
	orbits are being established, and (b) shows the absolute value of $z$ over 1 second.
	
	\subsection{Degenerate Hopf bifurcation}
	\label{sec:degenerate_hopf}
	
	\Cref{sec:derivation_of_hopf} introduced the first Lyapunov coefficient and its 
	effects on the stability or instability of the solutions. If the first Lyapunov 
	coefficient vanishes, $b=0$, a degenerate bifurcation occurs (also known as a Bautin 
	or generalised bifurcation).
	\begin{equation}
	\dot{z} = (\mu_0 + j\omega_0)z, \; z \in \mathbb{C}
	\label{eq:bautin}
	\end{equation}
	
	The non-linear term disappears, but there are still periodic orbits, as the limit 
	cycle `degenerates' into the plane at $\mu=0$ in $(x_1, x_2, \mu)$-space.
	This creates the relatively simple equation
	\begin{equation}
	\dot{z} = j \omega_0 z
	\label{eq:bautin_simple}
	\end{equation}
	for a degenerate bifurcation exhibiting a periodic solution and 
	\begin{equation}
	\dot{z} = j \omega_0 z + F(t)
	\label{eq:bautin_simple_forced}
	\end{equation}
	for a forced bifurcation.

	\subsection{Implementation of the Hopf bifurcation in software}
	\label{sec:implementation}
	
	By constructing a bank of resonators (`detectors'), each operating at a Hopf bifurcation and 
	tuned to a given characteristic frequency, then driving this with the input signal, we can 
	obtain  the $|z|$ responses of each detector.
	Frequency components in the input can be detected by analysing this output to 
	find periods when a response is increasing in a manner similar to that in Figure 
	\ref{fig:forcing_example_100Hz_line}.
	
	As Equation \eqref{eq:hopf_equation_forced} cannot be discretized, numerical 
	approximations are used in the implementation. 
	The software allows users to choose between the central difference approximation and 
	the fourth order Runge-Kutta method.
	However, when the detectors are operating at a degenerate Hopf bifurcation, these 
	numerical approximations introduce errors, where, at high frequencies, the detector 
	frequency and the input frequency can appear to be mismatched.
	Adjusting the characteristic frequency of the detector can mitigate the effects of 
	this. For example, when the fourth order Runge-Kutta method is selected, a detector 
	nominally 
	operating at $2\kHz$ requires its frequency to be shifted by $0.066\%$; a $3\kHz$
	detector requires frequency adjustment of 0.25\%.
	A search normalisation facility automatically adjusts the characteristic 
	frequencies of the detectors, thus allowing the software to be used over a wider 
	range of frequencies.
	
	%
	%

	\section{Empirical investigation of DetectorBank characteristics}
	\label{sec:investigation}
	
	For this software to be useful, aspects of its response must be known, including the 
	bandwidth and time response of a single detector, the maximum bandwidth of the 
	DetectorBank and the effect of varying the forcing amplitude. It is also desirable to 
	determine how a detector behaves when its characteristic frequency is presented in 
	the presence of others.
	
	The following investigations are presented with musical applications in mind, so 
	typical audio sample rates are used and the frequency range of interest covers the 
	range of fundamental frequencies in music ($27.5\Hz$ to approximately $4.2\kHz$) and 
	the range of 
	human hearing ($20\Hz$ to $20\kHz$).
	
	The system is positioned at the bifurcation point. Unless 
	otherwise stated, the first Lyapunov coefficient is zero so the detectors 
	are operating at a degenerate Hopf bifurcation and a damping factor of $1\cdot 
	10^{-4}$ and forcing amplitude $X=5$ are used. These values produce an output which 
	falls within a range close to $|z| \leq 1$. As will be discussed in  
	\Cref{sec:lyapunov,sec:damping,sec:output_amplitude}, the 
	system turns out to be well conditioned when these parameters are altered.
	
	Both the fourth order Runge-Kutta and central difference methods are used, with and 
	without normalisation, over a range of frequencies and with a range of sample rates 
	($48\kHz$, $96\kHz$ and $192\kHz$).
	The test audio comprises generated sine waves at various frequencies. This has the 
	advantage that the precise frequencies are known, so a poor response for a certain 
	frequency can be attributed to the detector, not the input; however, a pure sine wave 
	has neither the complexity nor the variation of musical audio.

	\subsection{Frequency response}
	\label{sec:graphs_and_description}
	%
	%
	
	\begin{figure*}
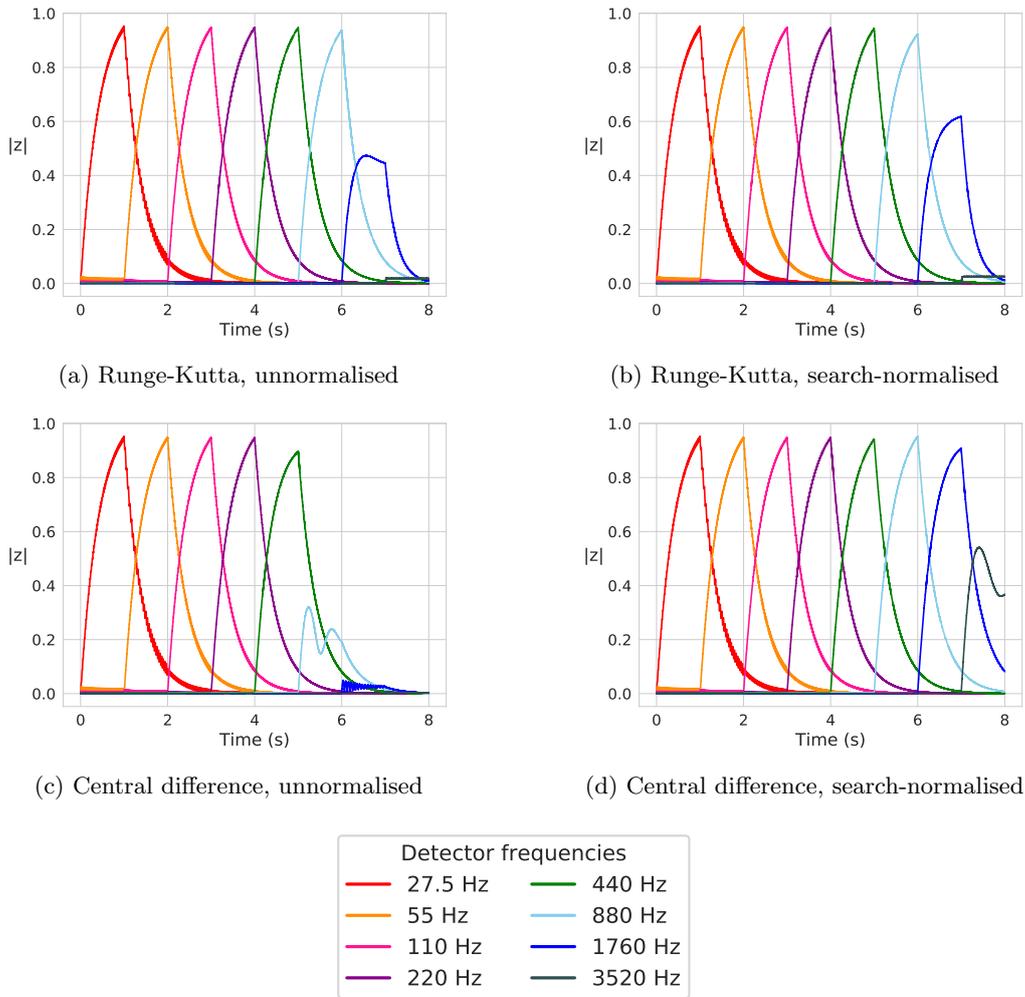

		\begin{subfigure}{0.5\textwidth}
			\centering
			\includegraphics[scale=\figscale]{sine_a_48_rk4_un.pdf}
			\caption{Runge-Kutta, unnormalised}
			\label{fig:sine_a_48}
		\end{subfigure}
		\begin{subfigure}{0.5\textwidth}
			\centering
			\includegraphics[scale=\figscale]{sine_a_48_rk4_sn.pdf}
			\caption{Runge-Kutta, search-normalised}
			\label{fig:sine_a_48_sn}
		\end{subfigure}
		
		\begin{subfigure}{0.5\textwidth}
			\centering
			\includegraphics[scale=\figscale]{sine_a_48_cd_un.pdf}
			\caption{Central difference, unnormalised}
			\label{fig:sine_a_48_cd}
		\end{subfigure}
		\begin{subfigure}{0.5\textwidth}
			\centering
			\includegraphics[scale=\figscale]{sine_a_48_cd_sn.pdf}
			\caption{Central difference, search-normalised}
			\label{fig:sine_a_48_cd_sn}
		\end{subfigure}
		
		\begin{subfigure}{\textwidth}
			\centering
			\includegraphics[scale=\figlegendscale]{a_legend.pdf}
		\end{subfigure}
		\caption{Responses to consecutive tones, each of which lasts for one second and increases 
		by two octaves, from $27.5\Hz$ to $1760\Hz$, with detectors tuned to each frequency. Both 
		numerical methods are tested, with and without frequency normalisation. $F_s=48\kHz$.}
		\label{fig:sine_a_48_all}
	\end{figure*}

	Figure \ref{fig:sine_a_48_all} shows the response to four consecutive tones, each 
	lasting for one second. The	frequencies ascend by two octaves from $27.5\Hz$ to 
	$1760\Hz$ (i.e. A0, A2, A4 and A6 in music). 
	The sample rate of the audio is $48\kHz$. It can be 
	seen that, for Runge-Kutta detectors, search-normalisation somewhat improves the 
	$1760\Hz$ response: the shape of the response is consistent with the lower frequencies, but the 
	amplitude is lower. 
	The $1760\Hz$ response of the central difference detectors is much improved by 
	search-normalisation. 
	Neither type of detector responds well to frequencies higher than this.

	\begin{figure*}
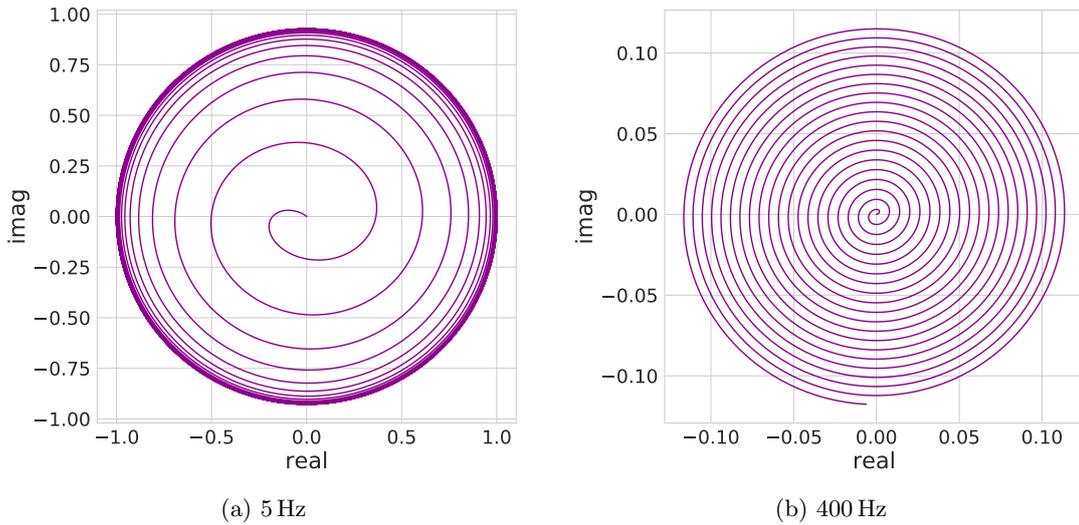

		\centering
		\begin{subfigure}{0.49\textwidth}
			\centering
            \includegraphics[scale=\ifcolour\smallfigscale\else\figscale\fi]{complex_response_5Hz_first20.pdf}
			\caption{$5\Hz$}
			\label{fig:complex_response_20Hz}
		\end{subfigure}
		\begin{subfigure}{0.49\textwidth}
			\centering
                \includegraphics[scale=\ifcolour\smallfigscale\else\figscale\fi]{complex_response_400Hz_first20.pdf}
			\caption{$400\Hz$}
			\label{fig:complex_response_400Hz}
		\end{subfigure}
		
		\caption{Complex response of the first 20 periods of two detectors at (a) $5\Hz$ 
			and (b) $400\Hz$. The elliptical nature of the orbits can be seen, particularly in 
			the lower frequency response.}
		\label{fig:complex_responses}
	\end{figure*}
		
\begin{figure}
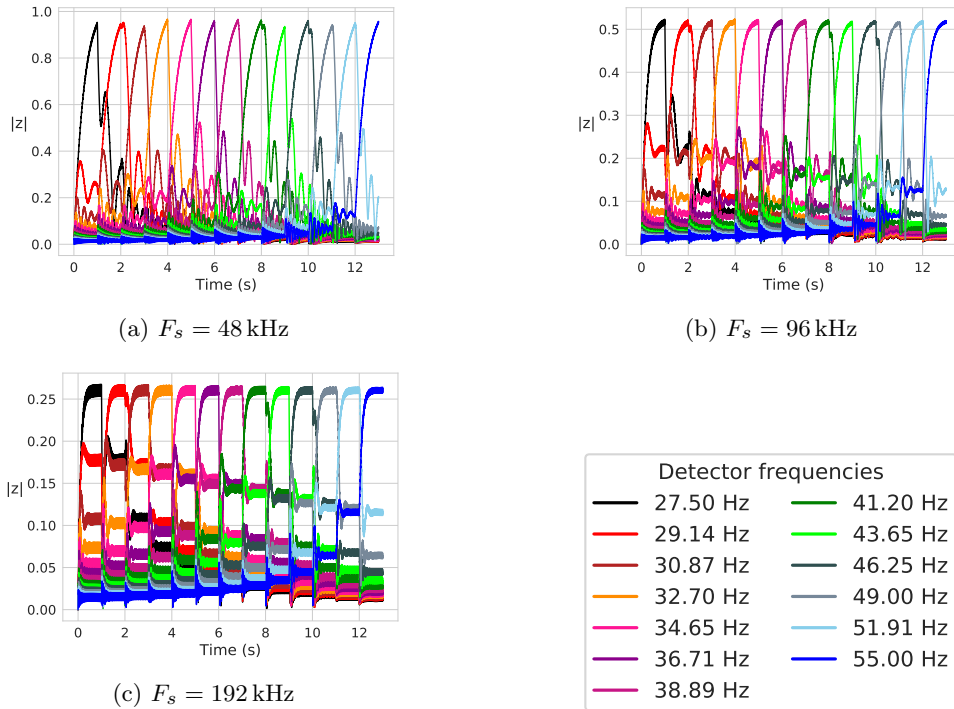

		\begin{subfigure}{0.49\textwidth}
			\centering
			\includegraphics[scale=\ifcolour\smallfigscale\else\figscale\fi]{sine_low_48_rk4_un.pdf}
			\caption{$F_s=48\kHz$}
			\label{fig:sine_low_48}
		\end{subfigure}
		\begin{subfigure}{0.49\textwidth}
			\centering
			\includegraphics[scale=\ifcolour\smallfigscale\else\figscale\fi]{sine_low_96_rk4_un.pdf}
			\caption{$F_s=96\kHz$}
			\label{fig:sine_low_96}
		\end{subfigure}
		
		\begin{subfigure}{0.49\textwidth}
			\centering
			\includegraphics[scale=\ifcolour\smallfigscale\else\figscale\fi]{sine_low_192_rk4_un.pdf}
			\caption{$F_s=192\kHz$}
			\label{fig:sine_low_192}
		\end{subfigure}
		\begin{subfigure}{0.49\textwidth}
			\centering
			\includegraphics[scale=\figlegendscale]{low_legend.pdf}
		\end{subfigure}
		\caption{Unnormalised Runge-Kutta response to sine tones generated at the two lowest 
		fundamental frequencies in the musical scale,
		$27.5\Hz$~(A0) and $29.1\Hz$~($A\sharp0$), presented consecutively.}
		\label{fig:sine_low_all}
	\end{figure}

	Small oscillations are noticeable in all the low frequency responses. These 
	oscillations are at twice the driving frequency and arise because
	the response in the complex plane is not quite circular, but slightly elliptical. 
	This eccentricity is greater at low frequencies than high, as can be seen in the graphs 
	presented in Figure \ref{fig:complex_responses}, which plot the first 20 periods of responses 
	at (a) $5\Hz$ and (b) $400\Hz$.
	Although increasing the sample rate increases the maximum frequency to which a detector will 
	respond, these oscillations in low frequency responses become even more pronounced, as 
	can be seen in Figure \ref{fig:sine_low_all}.
	Ultimately, this orbital eccentricity is corrected by amplitude normalisation, a feature that 
	will be explored in Section \ref{sec:output_amplitude}.

	Figure \ref{fig:sine_low_all} shows low frequency responses at the three sample 
	rates. The graphs show only the Runge-Kutta detector outputs, as there was no 
	significant difference between this and central difference. No frequency normalisation was 
	used. The detectors are being driven by two consecutive sine tones, generated at the 
	fundamental frequencies of the two lowest notes in the musical scale, A0 ($27.5\Hz$) and 
	$A\sharp0$~($29.1\Hz$).
	The responses at lower sample rates are more desirable, in that, although they have a 
	slower rate of response, they have better rejection of neighbouring semitones.
	For example, at all sample rates, when the $27.5\Hz$ tone sounds, the $29.1\Hz$ 
	detector also reacts briefly; however, the maximum amplitude of the $29.1\Hz$ detector 
	is $8.5\dB$ lower than that of the $27.5\Hz$ detector when the sample rate is 
	$48\kHz$ and only $2.3\dB$ lower at $192\kHz$.
	
	Again, looking at the responses when the $27.5\Hz$ tone is sounding, the rejection of 
	neighbouring frequencies happens very quickly. The $29.1\Hz$ detector begins to react 
	at the same time as the $27.5\Hz$ one, but reaches a maximum and stops 
	responding.
	At the three sample rates tested, $48\kHz$, $96\kHz$ and $192\kHz$, this difference of 
	$1.6\Hz$ is being discriminated in $265\ms$, $247\ms$ and $230\ms$, respectively.
	Therefore, here $\Delta t \Delta f = 0.464$, $0.404$ and $0.375$, all less than the lower limit 
	given by the uncertainty relation.
	
	When adjacent detectors respond, they 
	oscillate at the difference between the driving frequency and the 
	detector frequency. This means we might expect the local maximum to occur at half the 
	period of oscillation (in the case of the 27.5 and $29.1\Hz$ detectors, this would be 
	312.5\,ms). In reality, a time shift is observed due to the initial rate of change of the 
	response. This will be discussed further in Section \ref{sec:propinquitous_frequencies}.

	No normalisation is necessary for Runge-Kutta detectors operating at a sample rate of 
	$192\kHz$: they respond well to all frequencies up to the limit of fundamental frequencies 
	found in music.
	When $F_s=96\kHz$, the shape of the responses of unnormalised 
	Runge-Kutta detectors becomes distorted above about $3\kHz$. With 
	search-normalisation, the range can be extended to $4\kHz$. For both normalised and 
	unnormalised detectors, the amplitude of the responses decreases as the frequency 
	increases.
	The distorted shape means the detector's characteristic frequency is not quite matched to 
	the input frequency; decreasing amplitude means the detector is not responding as 
	strongly, although the frequency may still be correct. The first of these problems 
	can be mitigated with frequency normalisation, as described in Section 
	\ref{sec:implementation}; the second may be solved by scaling the responses.

	\subsection{Amplitude scaling}
	\label{sec:amp_scaling}
	
	
	As seen in Section \ref{sec:graphs_and_description}, as the characteristic frequency 
	and input frequency of a detector are increased, the amplitude of the responses 
	decreases. 
	Despite this amplitude decay, the shape of the responses is retained; therefore this 
	is not the result of distorted frequencies: the detector response is weaker, but 
	equally sharp, at higher frequencies.
	
	Maximum response values for various characteristic frequencies were found for all 
	detector types at a sample rate of $48\kHz$. 
	Using these values to scale up the detector responses creates a consistent output over a larger 
	range of frequencies.

	\subsection{System bandwidth}
	\label{sec:system_bandwidth}
	
	The Nyquist rate is a fundamental principle of linear signal processing. It states 
	that the sampling frequency must always be at least twice the bandwidth of the signal 
	you wish to represent. 
	For this non-linear system, a uniform output can only be obtained up to a few 
	kilohertz, even when amplitude scaling is employed. This suggests that,
	in a practical implementatation, the sample 
	rate must be several times the maximum frequency.
	
	Of course, the bandwidth of a signal is the difference between the maximum and 
	minimum frequencies, where the minimum need not be zero. Indeed, the figures in  
	Section \ref{sec:graphs_and_description} show that, to analyse the full music 
	fundamental frequency range with a single choice of method, normalisation and sample 
	rate, there will always be a small amount of deterioration at very low or very high 
	frequencies. It is also more computationally efficient to use a lower sample rate, 
	although this substantially reduces the maximum frequency.
	The question then arises: can the DetectorBank be made to cover the full range of 
	frequencies given by the Nyquist rate, without resorting to unreasonably high sample 
	rates?
	
	One possibility is frequency shifting of the signal to a more appropriate range. 
	In our system, this is achieved by generating a double sideband  
	signal, then subtracting a quadrature phase shifted version of the signal, which 
	leaves only the upper sideband \cite{2001van_modulation_theory}. 
	A phase shift can be implemented using the Hilbert 
	transform \cite{1975rabiner_dsp}.

	\begin{figure}
		\centering
		\includegraphics[width=0.8\textwidth]{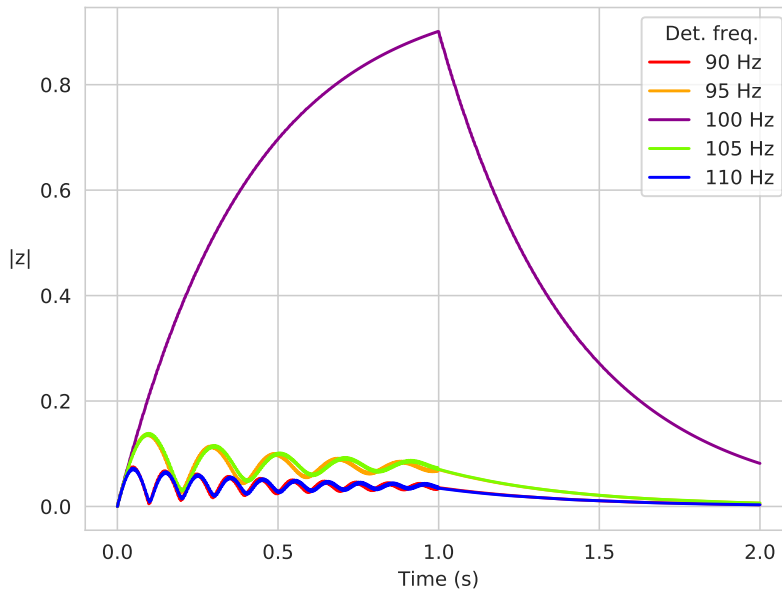}
		\caption{Responses of unnormalised Runge-Kutta detectors to a tone which was 
			generated at $4\kHz$ and shifted down to $100\Hz$.}
		\label{fig:ssb_hopf_output}
	\end{figure}
	

	Figure \ref{fig:ssb_hopf_output} shows the response of an unnormalised Runge-Kutta 
	detector, tuned $100\Hz$, and driven by a sine wave which was 
	generated at $4\kHz$ --- far beyond the empirically determined maximum for an 
	unnormalised Runge-Kutta detector --- and shifted down to $100\Hz$. The tone lasts 
	for one second and is followed by one second of silence. The sample rate is 
	$48\kHz$. Figure \ref{fig:ssb_hopf_output} also shows the responses of detectors at $\pm5\Hz$ 
	and $\pm10\Hz$. Frequency shifting does not appear to have a deleterious effect on the 
	responses.
	
	
	This process can also be used to shift the frequencies up. This will generally not be 
	required in musical applications, as detector responses for $27.5\Hz$ and greater are 
	adequate when the sample rate is $48\kHz$, but for other applications it may be 
	useful. 
	
	\begin{table}
		\centering
		\caption{Empirically determined frequency thresholds above which frequency shifting will
			be applied for each combination of numerical method (fourth order 
			Runge-Kutta, RK4, and central difference, CD) and normalisation 
			(search normalisation or unnormalised).}
		\label{tab:mod_f}
		\begin{tabular}{|c|c|c|}
			\hline
			\textbf{Method} & \textbf{Normalisation} & \textbf{$f_t$ (Hz)} \\
			\hline
			RK4 & none & 1600\\ \hline
			RK4 & search & 2200\\ \hline
			CD & none & 500\\ \hline
			CD & search & 700\\ \hline		
		\end{tabular}
	\end{table}
	
	When running the software, frequency shifted versions of the input will automatically
	be generated if any of the requested detector frequencies are above a certain
	threshold, $f_t$, given by the numerical methods and normalisation 
	(see Table \ref{tab:mod_f}). Frequencies will be shifted into the range $50\Hz$ to 
	$f_t+50\Hz$ to guarantee a satisfactory response.
	
	\begin{figure}
		\centering
		\ifcolour
		\includegraphics[width=0.8\textwidth]{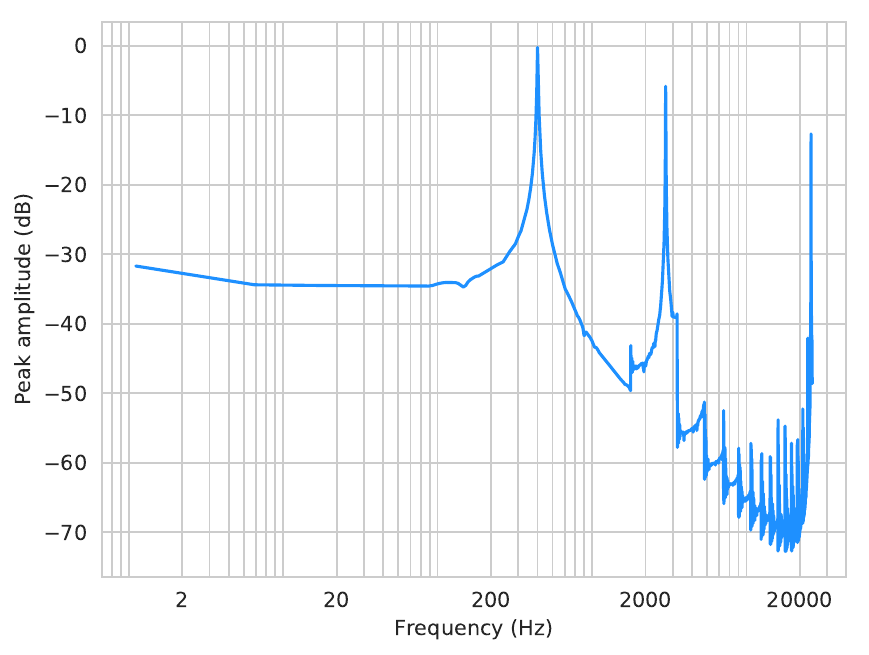}
		\else
		\includegraphics[width=0.8\textwidth]{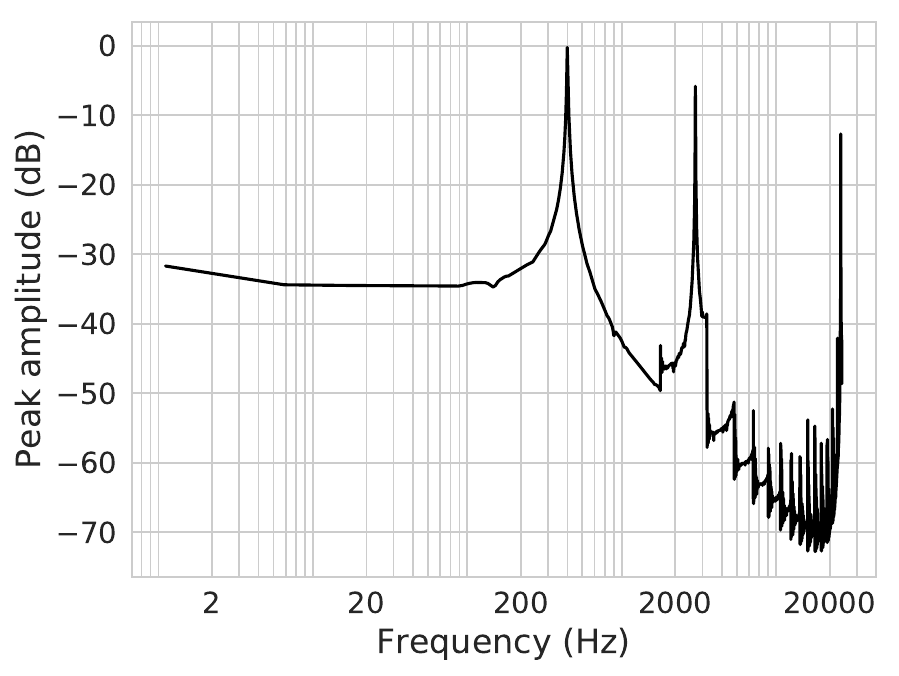}
		\fi
		\caption{DetectorBank response to $400\Hz$ tone}
		\label{fig:freq_response}
	\end{figure}

	Figure \ref{fig:freq_response} shows the frequency response of a DetectorBank, using the 
	Runge-Kutta method, with detectors tuned to frequencies up to the Nyquist rate, when presented 
	with a $400\Hz$ tone. As well as the expected peak at $400\Hz$, there are two lower peaks 
	at $23.6\kHz$ and $2.7\kHz$. The former may be due to the non-linearity of the system; as peaks 
	at $F_s/2 - f_{\textrm{in}}\Hz$ will generally be beyond the range of frequencies of interest 
	in musical 
	applications --- and as their scale, extent and location are known --- they will not be 
	investigated further here.
	
	The peak at $2.7\kHz$ is an artefact of frequency shifting. For unnormalised Runge-Kutta 
	detectors, frequencies above $1.6\kHz$ will be shifted into the range $50$--$1650\Hz$, i.e.  
	frequencies $n$ times above $1.6\kHz$ will be shifted down by $1600(n-1)+1550\Hz$. Peaks then 
	appear at double $1550\Hz$ minus the input frequency. Employing a better method of frequency 
	shifting may solve this problem, but at the expense of increased latency.
	
	From here on, unless otherwise stated, the sample rate used to characterise detector 
	responses will be $48\kHz$, as frequency shifting, amplitude scaling and search 
	normalisation result in responses that are sufficient for musical applications.

	\subsection{Propinquitous frequencies}
	\label{sec:propinquitous_frequencies}

	\begin{figure}
		\centering
		\ifcolour
		\includegraphics[width=0.8\textwidth]{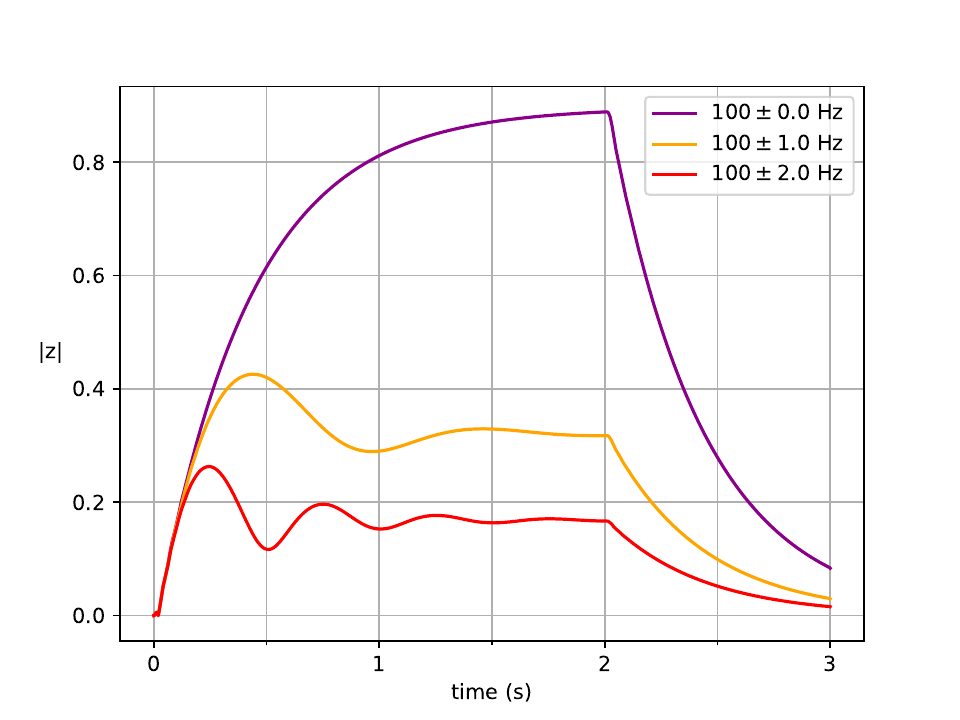}
		\else
		\includegraphics[width=0.8\textwidth]{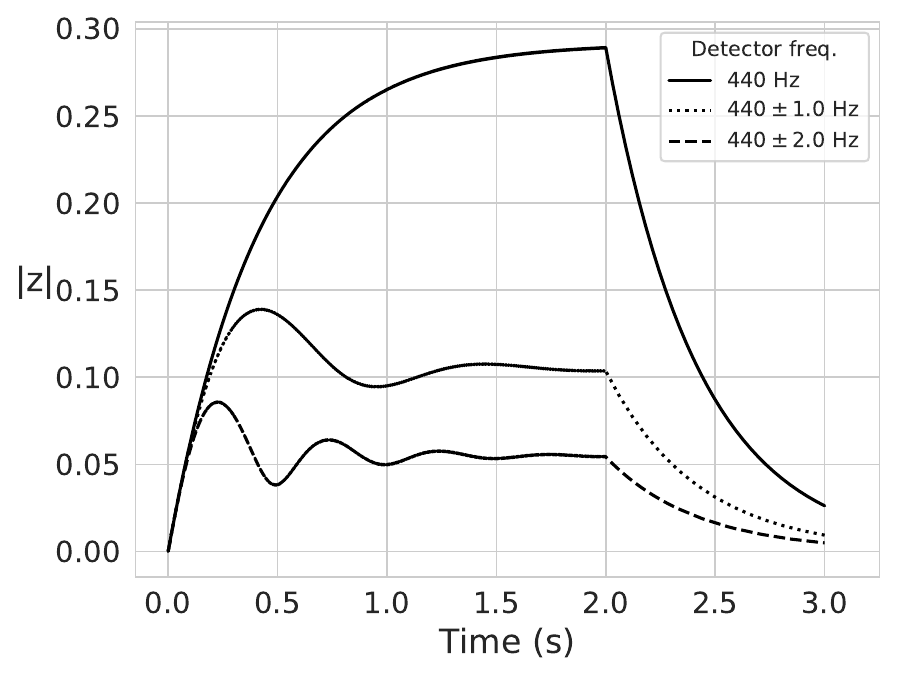}
		\fi 
		\caption{Responses of degenerate detectors at the driving frequency ($f_{in}=440\Hz$), 
			$f_{in}\pm1\Hz$ and $f_{in}\pm2\Hz$.}
		\label{fig:pm_1_2_Hz}
	\end{figure}

	
	A degenerate detector still responds when the driving frequency is close to 
	characteristic frequency.
	As the discrepancy between the frequencies increases, the response becomes weaker.
	The range of frequencies to which it will respond, and the strength with which it responds, are 
	determined by the bandwidth of the detector.

	
	
	When driven off-centre, the output will also oscillate at a rate of the difference between the 
	driving and characteristic frequencies.
	Therefore, it may be expected that the oscillation would reach its peak after half a period; 
	however, it invariably takes a shorter time than this.
	For degenerate detectors operating at $0.5\Hz$ from the driving frequency (and with a 
	damping of $1 \cdot 10^{-4}$) the time advance is about $230\ms$, which decreases to 
	$4\ms$ when the frequency difference increases to $5\Hz$.
	
	This time shifting is due to different parameters dominating the response in the transient and 
	steady state.
	Immediately after forcing begins, all responses react at the same rate, 	
	determined by the first Lyapunov coefficient, damping factor and sample rate. 
	After a certain time (approximately $100\ms$ for the detectors shown in 
	Figure \ref{fig:pm_1_2_Hz}) the responses reach a steady state, where the main 
	factor regulating the response shape is the similarity between the detector and 
	driving frequencies.
	
	As the oscillation frequency may be ascertained by measuring the time from maximum to 
	the following local minimum, this time shifting would allow $\Delta f$ to be obtained 
	sooner. 
	
	\subsection{Detector bandwidth}
	
	%
	
	\label{sec:lyapunov}
	Any algorithm which analyses the responses may miss events where the detector and 
	driving frequencies do not quite match. For example, as can be seen in Figure 
	\ref{fig:pm_1_2_Hz}, the maximum response for detectors at $\pm 1\Hz$ from the  driving 
	frequency is roughly half that of the correct 
	detector.
	In fact, the ratio of maximum amplitude of the matched response to that of a slightly 
	mismatched response is approximately:
	\begin{equation}
	\label{eq:max(z)_ratio}
	\frac{\max (\rvert z_d\rvert)}{\max (\rvert z_0\rvert)}
        \approx
        \frac{1}{\rvert f_{\mathrm{in}} - f_0\rvert + 1}
	\end{equation}
	where $z_d$ is the mismatched detector response, $z_0$ is the matched detector 
	response, $f_{\mathrm{in}}$ is the driving frequency and $f_0$ is the detector 
	frequency.\footnote{This only holds when the sample rate is $48\kHz$. At higher 
		sample 
		rates, the maximum values of propinquitous detectors are higher.}
	This suggests that the $3\dB$ point of a detector is about 
	$\pm0.41\Hz$ from the detector's frequency. Experimental measurements put 
	the $3\dB$ point at $\pm0.58\Hz$, i.e. the detector bandwidth is 
	$1.16\Hz$ (when the sample rate is $48\kHz$, the damping is $1\cdot 10^{-4}$ and the 
	first Lyapunov coefficient is zero).
	
	This extremely sharp cutoff will be good for many applications, but for some --- e.g. 
	music analysis, where performed notes will rarely be at exactly the `correct' 
	frequency, but will be perceived as being the correct pitch --- it may be desirable 
	to widen the response of the detectors. This can be achieved by using a 
	non-degenerate Hopf bifurcation, i.e. one where the first Lyapunov coefficient,~$b$, 
	is not zero. For a supercritical bifurcation (stable periodic solutions), $b$ should 
	be negative.
	
	The value of $b$ required for a $3\dB$ point at $\pm1\Hz$ from 
	the detector frequency for a purely sinusoidal input with amplitude $X=25$ was 
	experimentally determined as 
	approximately $-0.16$, which rapidly increased as the frequency difference (in Hertz) 
	is increased (see table \ref{tab:lyapunov_3dB_pmHz}).
	
	\begin{table}
		\centering
		\caption{Empirical values for First Lyapunov coefficient ($b$) required for a 
			$-3\dB$ point at $\pm$ 1 to $5\Hz$, i.e. a detector bandwidth of 2 
			to $10\Hz$.}
		\label{tab:lyapunov_3dB_pmHz}
		\begin{tabular}{|c|c|}
			\hline
			\textbf{\begin{tabular}[c]{@{}c@{}}Bandwidth\\(Hz)\end{tabular}} & 
			\textbf{\begin{tabular}[c]{@{}c@{}}First Lyapunov \\ 
					coefficient\end{tabular}} \\ 
			\hline
			$2$ & $-0.160$   \\ \hline	
			$4$ & $-1.278$   \\ \hline	
			$6$ & $-4.303$   \\ \hline	
			$8$ & $-10.183$  \\ \hline	
			$10$ & $-19.863$ \\ \hline		
		\end{tabular}
	\end{table}
		
	A proportionality is observed between the log of the values in Table 
	\ref{tab:lyapunov_3dB_pmHz}. Genetic algorithms (implemented using DEAP 
	\cite{2012fortin_deap}) were used to confirm this relationship for bandwidths up to 
	$100\Hz$. 
	From these values, an equation relating bandwidth, $B$, and first Lyapunov 
	coefficient, $b$, can be derived:
	\begin{equation}
	b = -\mathrm{exp}\Big(m \big(\ln(B) - \ln(x_0)\big) + \ln(y_0)\Big)
	\label{eq:lyapunov_bandwidth_full}
	\end{equation}
	where $x_0=2$, $x_1=10$, $y_0=0.15$, $y_1=20$ and 
	$m = \big(\ln(y_1)-\ln(y_0)\big)/\big(\ln(x_1)-\ln(x_0)\big) \approx 3$.
	This can then be simplified to 
	\begin{equation}
	b = -0.02 B^3
	\label{eq:lyapunov_bandwidth}
	\end{equation}
	
	\begin{table}
		\centering
		\caption{Empirical values for First Lyapunov coefficient required for a 
			$3\dB$ point at bandwidths of 2--$10\Hz$ at high sample rates, with $X=25$.}
		\label{tab:lyapunov_3dB_pmHz_high_sr}
		\begin{tabular}{|c|c|c|c|}
			\hline
			\textbf{\begin{tabular}[c]{@{}c@{}}Bandwidth\\(Hz)\end{tabular}} & 
			\textbf{\begin{tabular}[c]{@{}c@{}}First Lyapunov \\ coefficient\\ 
					$f_s=96\kHz$\end{tabular}} & \textbf{\begin{tabular}[c]{@{}c@{}}First Lyapunov 
					\\ 
					coefficient\\ $f_s=192\kHz$\end{tabular}} \\ \hline
			2 & $-0.04$ & ---       \\ \hline
			4 & $-1.25$ & $-0.32$   \\ \hline
			6 & $-4.44$ & $-3.60$   \\ \hline
			8 & $-10.64$ & $-9.99$   \\ \hline
			10 & $-20.61$ & $-20.43$ \\ \hline
		\end{tabular}
	\end{table}
	
	The first Lyapunov coefficients required for bandwidths of 2--$10\Hz$ at higher 
	sample 
	rates 
	($96\kHz$ and $192\kHz$) did not display the same trend (see Table 
	\ref{tab:lyapunov_3dB_pmHz_high_sr}). 
	As the sample rate is increased, the minimum detector bandwidth increases and so 
	smaller values of first Lyapunov coefficient are required to widen the response.
	For $192\kHz$, there is no value of $b$ small enough to bring the maximum 
	responses of detectors at $\pm 1\Hz$ down to $-3\dB$. The limit 
	seems to be around $-1.36\dB$. However, for larger bandwidths, e.g. 8 or $10\Hz$, 
	the first Lyapunov coefficient required is very similar at all tested sample rates.
	
	
	As the first Lyapunov coefficient is increased, the magnitude of all the responses 
	decreases, even those where the input frequency and the detector frequency are 
	exactly matched. 
	This problem is addressed by amplitude normalisation, discussed in Section 
	\ref{sec:output_amplitude}.
	

	\subsubsection{Effect of varying forcing amplitude}
	\label{sec:forcing_amplitude}	
	
	Increasing the forcing amplitude widens the bandwidth of a detector.
	The relationship between bandwidth, $B$, and first Lyapunov coefficient, $b$, given in 
	Equation \eqref{eq:lyapunov_bandwidth} was experimentally determined for 
	sinusoidal forcing with a constant amplitude $X=25$. It is therefore necessary to 
	adapt this to accommodate an arbitrary forcing amplitude.
	
	
	The relationship between forcing amplitude and first Lyapunov coefficient was derived 
	empirically with genetic algorithms.
	Using this data, the following function was found to
	scale the values returned by Equation \eqref{eq:lyapunov_bandwidth} (denoted $b_0$, with input 
	amplitude $X_0=25$) to give a $3\dB$ point at the desired bandwidth for an arbitrary input 
	amplitude $X_1$
	\begin{equation}
	b_1 =  b_0 \bigg(\frac{X_0}{X_1}\bigg)^2
	\label{eq:lyapunov_amplitude_scaling}
	\end{equation}
	
	Therefore, the first Lyapunov coefficient required to create a detector with a bandwidth $B$ 
	for a forcing amplitude $X$ is
	\begin{equation}
	b = - \frac{12.5 B^3}{X^2}
	\label{eq:lyapunov_bandwidth_amplitude}
	\end{equation}
	
	This relationship between forcing amplitude, first Lyapunov coefficient and bandwidth 
	is constant for damping factors up to $5 \cdot 10^{-4}$, except for some variation 
	for smaller bandwidths (up to about $10\Hz$), as the damping factor affects the 
	minimum bandwidth.	
	
	\subsection{Damping}
	\label{sec:damping}
	
	\begin{figure}
		\centering
		\includegraphics[width=0.8\textwidth]{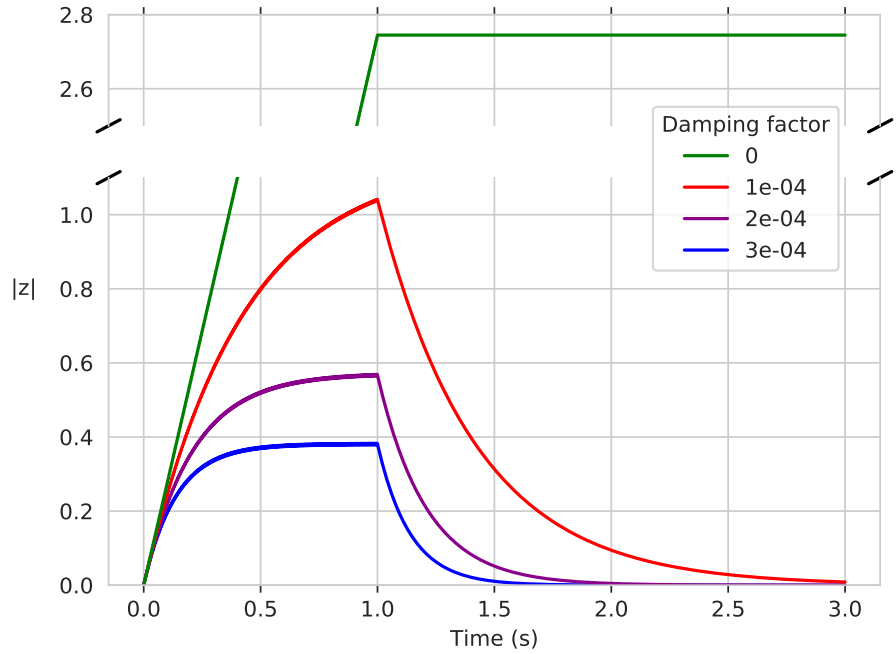}
		\caption{Responses of damped and undamped detectors.}
		\label{fig:damping}
	\end{figure}
	
	Figure \ref{fig:damping} shows the responses of four detectors to a one second sine 
	tone. One detector is undamped; this response increases at a uniform rate, but does not
	return to zero after the tone stops. The other three detectors have	
	different damping levels: $1\cdot 10^{-4}$, $2\cdot 10^{-4}$ and $3\cdot 10^{-4}$ 
	(the first of these is the damping factor which has been used in all the experiments here 
	thus far). 
	As would be expected, as the damping factor increases, the maximum amplitude becomes 
	lower and the bandwidth becomes wider.
	
	Minimum detector bandwidths were experimentally determined for five damping levels from 
	$1\cdot 10^{-4}$ to $5\cdot 10^{-4}$ by presenting a tone at $440\Hz$ to a number of 
	detectors around this input frequency at a fraction of a Hertz increments. 
	The bandwidth was then determined by finding the frequencies at which the response magnitudes 
	were $3\dB$ lower than that of the centre.
	
	The first Lyapunov coefficients returned by Equation \eqref{eq:lyapunov_bandwidth_amplitude} 
	were compared with experimentally determined values for a range of bandwidths above these 
	minima.
	The relationship given in Equation \eqref{eq:lyapunov_bandwidth_amplitude} was 
	found to give accurate values, except for bandwidths close to the minimum at the highest 
	damping factor. The minimum bandwidth that can be represented accurately with a damping factor 
	of $5\cdot 10^{-4}$ has, therefore, been increased from $4.56\Hz$ to $4.86\Hz$. 
	Table \ref{tab:damping_bandwidth} gives all the minimum bandwidths for different damping 
	factors and sample rates.
	
	\begin{table}[t]
		\centering
		\caption{Minimum bandwidths for different levels of damping and sample rates.}
		\label{tab:damping_bandwidth}
		\begin{tabular}{|c|c|c|c|}
			\hline
			\textbf{Damping} & \textbf{\begin{tabular}[c]{@{}c@{}}Bandwidth (Hz)\\ 
					$f_s=48\kHz$\end{tabular}} & \textbf{\begin{tabular}[c]{@{}c@{}}B'width (Hz)\\ 
					$f_s=96\kHz$\end{tabular}} & \textbf{\begin{tabular}[c]{@{}c@{}}B'width (Hz)\\ 
					$f_s=192\kHz$\end{tabular}} \\ \hline
			{$1 \cdot 10^{-4}$} & 0.922 & 1.824 & 3.653  \\   \hline
			{$2\cdot 10^{-4}$}  & 1.832 & 3.648 & 7.307  \\   \hline
			{$3\cdot 10^{-4}$}  & 2.752 & 5.496 & 11.040 \\   \hline
			{$4\cdot 10^{-4}$}  & 3.606 & 7.328 & 14.700  \\  \hline
			{$5\cdot 10^{-4}$}  & 4.860  & 9.160  & 18.367 \\ \hline
		\end{tabular}
	\end{table}
	
	From these values it can be seen that the detector bandwidth changes linearly with 
	damping at all three sample rates.
	

%

	\begin{table}[t]
		\centering
		\caption{Rise times as damping factor is changed, $F_s=48\kHz$}
		\label{tab:rise_times}
		\begin{tabular}{|c|c|c|c|}
			\hline
		\textbf{Damping}           & \textbf{10\% time (ms)} & \textbf{90\% time (ms)} & 
		\textbf{Rise time (ms)} \\ \hline
		\textbf{$1 \cdot 10^{-4}$} & 43.8125 & 956.438 & 912.625 \\ \hline
		\textbf{$2 \cdot 10^{-4}$} & 22.0833 & 480.229 & 458.146 \\ \hline
		\textbf{$3 \cdot 10^{-4}$} & 14.4583 & 320.042 & 305.583 \\ \hline
		\textbf{$4 \cdot 10^{-4}$} & 10.8958 & 240.458 & 229.562 \\ \hline
		\textbf{$5 \cdot 10^{-4}$} & 8.66667 & 191.729 & 183.062 \\ \hline
		\end{tabular}
		
		\vspace{10pt}
		
		\caption{Relaxation times as damping factor is changed, $F_s=48\kHz$}
		\label{tab:relaxation_times}
		\begin{tabular}{|c|c|}
			\hline
			\textbf{Damping}           & \textbf{Relaxation time (ms)} \\ \hline
			\textbf{$1 \cdot 10^{-4}$} & 416.396                       \\ \hline
			\textbf{$2 \cdot 10^{-4}$} & 208.104                       \\ \hline
			\textbf{$3 \cdot 10^{-4}$} & 138.667                       \\ \hline
			\textbf{$4 \cdot 10^{-4}$} & 103.958                       \\ \hline
			\textbf{$5 \cdot 10^{-4}$} & 83.1250                       \\ \hline
		\end{tabular}
	\end{table}
	
	\Cref{tab:rise_times,tab:relaxation_times} give the rise and relaxation times at different 
	damping factors, where the rise time is the time taken to go from $10\%$--$90\%$ of the maximum 
	amplitude and the relaxation time is the time taken for the amplitude of the response to fall 
	to 1/e of the maximum.
	The rise and relaxation times are both inversely proportional to the damping.
		
	Taken together, \Cref{tab:damping_bandwidth,tab:rise_times,tab:relaxation_times} show 
	that there is a trade-off between improving time performance of the detectors and 
	reducing the frequency selectivity. 
	
%
	
	
	Increasing the damping has a similar effect to increasing the sample rate (see Section 
	\ref{sec:lyapunov}): the minimum bandwidth of a detector is also increased.
	This similarity arises because, when implemented in software, each $z$ value is scaled by both 
	$1/\mathrm{sr}$ and $1-d$.
	

	\subsection{Output amplitude}
	\label{sec:output_amplitude}
	
	In non-linear systems, the principle of superposition does not apply, so all input 
	parameters must be considered when describing the output of the system.
	
	Amplitude scaling (see Section \ref{sec:amp_scaling}) corrects any aberration in 
	output amplitude due to characteristic frequency, numerical method or frequency 
	normalisation. However, other DetectorBank parameters --- damping factor, sample 
	rate, forcing amplitude and detector bandwidth can affect the output amplitude. 
	
	
	
	Amplitude normalisation can be used to keep the DetectorBank output within the range 
	$-1$ to 1, allowing comparison and analysis of the output of detectors with different 
	bandwidths or DetectorBanks with different parameters.
	
	\begin{figure*}
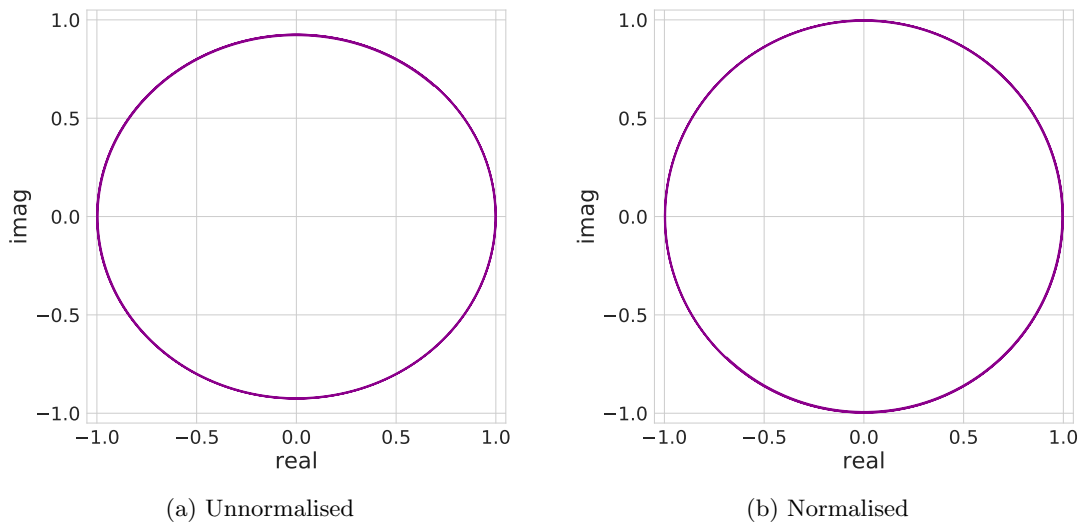

		\centering
		\begin{subfigure}{0.49\textwidth}
			\centering
			\includegraphics[scale=\ifcolour\smallfigscale\else\figscale\fi]{complex_response_5Hz_last5_again.pdf}
			\caption{Unnormalised}
			\label{fig:complex_response_5Hz_last_5_unnorm}
		\end{subfigure}
		\begin{subfigure}{0.49\textwidth}
			\centering
			\includegraphics[scale=\ifcolour\smallfigscale\else\figscale\fi]{complex_response_5Hz_last5_norm.pdf}
			\caption{Normalised}
			\label{fig:complex_response_5Hz_last_5_norm}
		\end{subfigure}
		
		\caption{The last five periods of a $5\Hz$ response in the complex plane, (a) without and 
			(b) with amplitude normalisation. It can be seen that amplitude normalisation corrects 
			the 
			orbital eccentricity.}
		\label{fig:complex_responses_norm}
	\end{figure*}
	
	It can also correct orbital eccentricity, which causes the small oscillations in the responses, 
	discussed in Section \ref{sec:graphs_and_description}.
	This is achieved by finding the ratio of the maximum real and imaginary parts and using this to 
	scale the imaginary part of the response.
	
	Figure \ref{fig:complex_responses_norm} shows the results of this. Figure 
	\ref{fig:complex_response_5Hz_last_5_unnorm} shows the last five periods of a $5\Hz$ response. 
	These are clearly elliptical and have an eccentricity of $7.475\cdot10^{-2}$. In Figure 
	\ref{fig:complex_response_5Hz_last_5_norm}, this response has been normalised and the 
	eccentricity has dropped by a factor of eight, to $9.347\cdot10^{-3}$.
	
	\subsection{Detector response in the presence of wideband noise}
	\label{sec:noise}
	
	Nonlinearities may introduce artefacts at frequencies within the range of interest. 
	There is, therefore, a legitimate cause for concern that this system will give false 
	results when multiple frequencies are presented simultaneously. 
	
	
	
	%
%
%
	
\begin{figure*}
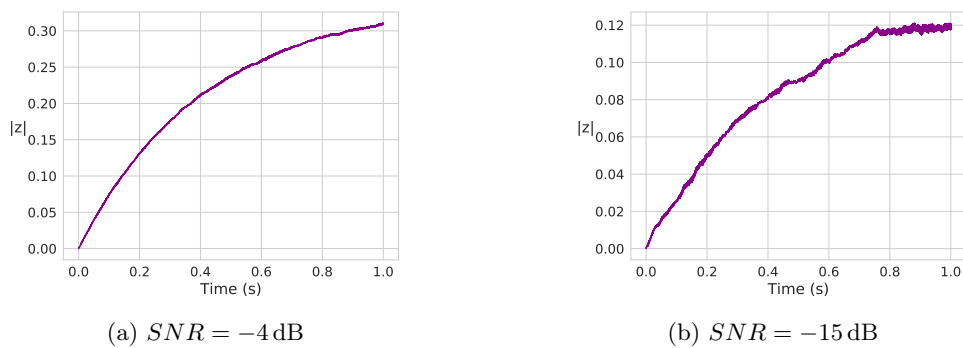

		\centering
		\begin{subfigure}{0.49\textwidth}
			\centering
			\includegraphics[scale=\ifcolour\smallfigscale\else\figscale\fi]{noise_and_tone_440Hz_snr-4dB.pdf}
			\caption{$SNR = -4\dB$}
			\label{fig:noise_tone_snr-4dB}
		\end{subfigure}
		\begin{subfigure}{0.49\textwidth}
			\centering
			\includegraphics[scale=\ifcolour\smallfigscale\else\figscale\fi]{noise_and_tone_440Hz_snr-15dB.pdf}
			\caption{$SNR = -15\dB$}
			\label{fig:noise_tone_snr-15dB}
		\end{subfigure}
		\caption{Responses to tones in the presence of white noise, where the amplitude of the 
		noise exceeds the tone by (a) $4\dB$ and (b) $15\dB$.}
		\label{fig:noise_tone_snr}
	\end{figure*}

	Figure~\ref{fig:noise_tone_snr} shows the response to white noise and a sine tone 
	presented simultaneously. 
	The signal-to-noise ratios (SNR) of these are (a) $-4\dB$ and (b) $-15\dB$. Although the 
	responses become somewhat distorted and the amplitude is decreased when noise is introduced, 
	the general shape of the response is preserved.
	
	Supplementary materials containing additional figures relating to
	this section may be obtained from \url{}.
	
	\section{Conclusions}
	\label{sec:conclusions}
	
	Empirical investigation determines that it is possible to obtain 
	precise information about time and frequency characteristics of a signal 
	simultaneously, beyond what is possible with Fourier analysis,
	by using non-linear tuned resonators (`detectors').
	
	The characteristics of the detectors depend on the input parameters, but it is 
	possible to construct detectors with a narrow bandwidth (down to $0.922\Hz$) which 
	will reject frequencies outwith this range in less than half the period of 
	oscillation. 
	
	The deficiencies brought about by the use of numerical approximations to 
	implement the Hopf equation can be circumvented with various techniques including 
	frequency shifting, frequency normalisation, amplitude scaling and amplitude 
	normalisation. Although these methods can introduce artefacts at high frequencies, 
	frequencies up to $1.6\kHz$ can be represented without recourse to shifting or normalisation. 
	This covers $80\%$ of the range of fundamental frequencies likely to appear in music.
	
	Potential problems associated with non-linear systems and multiple simultaneous input 
	frequencies do not emerge here: the system can withstand a large amount of noise in 
	the input signal, producing useful results when the noise exceeds the signal by $15\dB$ or more.
	
	For optimal output in any given situation, a number of parameters must be 
	considered --- the first Lyapunov coefficient (and its relationship to bandwidth), sample 
	rate, damping, forcing amplitude --- and must be weighed against the desired 
	performance --- time response, maximum amplitude and smoothness of response.
	
	The software presented here	forms a suitable basis for musicological applications,
	like a note onset detector which is required to detect 
	both note onset times and the frequency components present.
	

\end{document}